\documentclass[12pt]{article}
\usepackage{epsf}

\begin{document}

\newcommand{\gtrsim}{ \mathop{}_{\textstyle \sim}^{\textstyle >} }
\newcommand{\lesssim}{ \mathop{}_{\textstyle \sim}^{\textstyle <} }

\renewcommand{\thefootnote}{\fnsymbol{footnote}}
\setcounter{footnote}{0}

\begin{titlepage}

\def\thefootnote{\fnsymbol{footnote}}

\begin{center}

\hfill TU-637\\
\hfill hep-ph/0110383\\
\hfill October, 2001\\

\vskip .5in

{\Large \bf
Muon Magnetic Dipole Moment and Higgs Mass\\
in Supersymmetric SU(5) Models
}

\vskip .45in

{\large
Motoi Endo and Takeo Moroi
}

\vskip .45in

{\em
Department of Physics,  Tohoku University, Sendai 980-8578, Japan
}

\end{center}

\vskip .4in

\begin{abstract}
    
    We study the muon magnetic dipole moment and the Higgs mass in the
    framework of the supersymmetric SU(5) models.  In this analysis,
    all the relevant parameters in the Lagrangian are taken to be
    free; in particular, assumption of the universal scalar mass is
    {\sl not} adopted.  Negative search for the Higgs boson at the LEP
    II experiment sets an important constraint on the supersymmetric
    contribution to the muon magnetic dipole moment $a_\mu({\rm
    SUSY})$.  It is shown that, for a fixed value of the lightest
    Higgs mass, the maximum possible value of $a_\mu({\rm SUSY})$
    becomes significantly larger in the general SU(5) case compared to
    the case of the universal scalar mass (i.e., the case of the
    so-called ``CMSSM'').  We also point out that, if we take
    relatively large value of the trilinear scalar couplings, the
    constraint from the Higgs mass is drastically relaxed.  In this
    case, $a_\mu({\rm SUSY})$ can be as large as $\sim 50\times
    10^{-10}$ even for small value of $\tan\beta$ (say, for
    $\tan\beta=5$).

\end{abstract}
\end{titlepage}

\renewcommand{\thepage}{\arabic{page}}
\setcounter{page}{1}
\renewcommand{\thefootnote}{\#\arabic{footnote}}
\setcounter{footnote}{0}

Currently, supersymmetry (SUSY) is regarded as one of the most
attractive candidates of the new physics beyond the standard model.
Most importantly, the minimal supersymmetric standard model (MSSM) is
not only consistent with experimental constraints but also is
suggested from precision measurements; precise measurements of the
electroweak parameters strongly prefer a light Higgs ($m_h\leq 205\ 
{\rm GeV}$ \cite{hph0102143}) and the MSSM naturally predicts such a
light Higgs boson.  In addition, it is well known that three gauge
coupling constants meet at the scale $M_{\rm GUT}\simeq 2\times
10^{16}\ {\rm GeV}$ if the renormalization group equations based on
the MSSM are used.  Thus, to construct a viable model of the grand
unified theory (GUT), it is natural to introduce superpartners of the
standard-model particles to realize the gauge coupling unification.
Furthermore, in supersymmetric models, the naturalness problem is
solved because of the cancellation of the quadratic divergences
between bosonic and fermionic loops.

As well as these, the Brookhaven E821 experiment provided a new
motivation of SUSY.  In February, 2001, the Brookhaven E821 experiment
reported their result on the precise measurement of the muon magnetic
dipole moment (MDM) \cite{PRL86-2227}:
\begin{eqnarray}
    a_{\mu}({\rm E821}) = 11\ 659\ 202\ (14)(6) \times 10^{-10}.
\end{eqnarray}
Comparing this value with the standard-model prediction
\cite{PRD64-013014}, we find $a_{\mu}({\rm E821})-a_{\mu}({\rm
SM})=43(16)\times 10^{-10}$, meaning that the E821 result is about
2.6-$\sigma$ away from the standard-model prediction.  If we take this
deviation seriously, some new physics beyond the standard model is
needed to explain this anomaly.  Among various models, the MSSM can
provide significant extra contribution to the muon MDM \cite{muonMDM}.

Of course, precise value of the SUSY contribution to the muon MDM
$a_{\mu}({\rm SUSY})$ depends on soft SUSY breaking parameters which
are model-dependent.  Therefore, it is important to study the SUSY
contribution to the muon MDM in various models to see if the E821
anomaly can be well explained without conflicting various experimental
constraints.  Indeed, after the announcement of the E821 result, there
have been many works which discussed the SUSY contribution to the muon
MDM in various cases \cite{recentMDM}.  In particular, with Komine and
Yamaguchi, one of the authors (T.M.) pointed out that, in the
unconstrained MSSM, $a_{\mu}({\rm SUSY})$ can be large enough to
explain the deviation in wide parameter region.

Since the GUT is a strong motivation to introduce SUSY, it is
reasonable to ask if the E821 anomaly can be explained even in the
framework of SUSY GUTs.  Once the unification of the gauge groups is
assumed, some of the coupling constants and mass parameters should
also obey the unification conditions, and hence the number of the free
parameters is reduced compared to the case of the unconstrained MSSM.
Thus, in SUSY GUTs, it is non-trivial whether the SUSY contribution to
the muon MDM can become large enough in parameter region consistent
with other experimental constraints.  Previously, in several works,
the SUSY contribution to the muon MDM is studied in SUSY SU(5) models.
In those works, however, a very strong assumption is adopted, that is,
the universal scalar mass at the unification scale.  (Such a model is
sometimes called ``constrained MSSM'' or ``CMSSM.'')  In general SUSY
SU(5) GUTs, however, the universal scalar mass is not necessarily
realized, and hence such an assumption imposes too strong constraints
on the model. Indeed, there are models which do not predict the
universal scalar mass.  In addition, even if the universal scalar mass
is somehow realized at the cutoff scale of the theory which naturally
is the gravitational scale, the renormalization group effect spoils
the universality.  Therefore, it is necessary to study the muon MDM in
a general framework of SUSY GUTs, and in this letter we consider the
SUSY contribution to the muon MDM in a general framework of the SUSY
SU(5) models.

Before going into the details, we make several comments.  First, in
order to eliminate the model-dependence as much as possible, we do not
consider constraints from flavor and CP violating processes which are
sensitive to small off-diagonal elements in the sfermion mass
matrices.  Such off-diagonal elements are hard to predict, and SUSY
contributions to flavor and CP violations are significantly affected
if the values of such off-diagonal elements are changed.  Second, we
do not take account of cosmological constraints.  In particular, we do
not require that the thermal relic of the lightest superparticle (LSP)
be the cold dark matter (CDM).  This is because the relic density of
the LSP depends on cosmological scenarios.  For example, if a
late-time entropy production exists, the relic density of the LSP is
changed \cite{NPB570-455}.  Furthermore, the LSP is not the only
particle-physics candidate of the CDM; for example, axion may be the
CDM.  As will be discussed, the model is still severely constrained by
the lightest Higgs mass $m_h$ even after excluding these constraints.
In the following discussion, we will see that the SUSY contribution to
the muon MDM is significantly constrained in some case once we impose
the constraint on the lightest Higgs mass.

We begin our discussion by introducing the model we consider.  We
study SUSY SU(5) models.  In this framework, the low-energy effective
theory below the GUT scale $M_{\rm GUT}$ is the MSSM which contains
the chiral superfields $Q_i({\bf 3}, {\bf 2}, 1/6)$, $U_i^c({\bf 3^*},
{\bf 1}, -2/3)$, $D^c_i({\bf 3^*}, {\bf 1}, 1/3)$, $L_i({\bf 1}, {\bf
2}, -1/2)$, $E^c_i({\bf 1}, {\bf 1}, 1)$, $H_u({\bf 1}, {\bf 2},
1/2)$, and $H_d({\bf 1}, {\bf 2}, -1/2)$ (where we denote the gauge
quantum numbers of the SU(3)$_C$, SU(2)$_L$, and U(1)$_Y$ gauge
interactions in the parentheses) as well as vector superfields
describing the SU(3)$_C$, SU(2)$_L$, and U(1)$_Y$ gauge multiplets.
Here, the subscript $i$ is the flavor index which runs from 1 to 3.
With these superfields, the relevant part of the superpotential is
given by\footnote
{For simplicity, we omit the gauge indices.  Our sign convention for
the $\mu_H$ parameter and the gaugino masses is the same as that used
in Ref.\ \cite{PRep117-75}.}
\begin{eqnarray}
    W_{{\rm MSSM}} &=& H_u U^c_i [Y_U]_{ij} Q_j
    + H_d D^c_i [Y_D]_{ij} Q_j
    + H_d E^c_i [Y_E]_{ij} L_j
    + \mu_H H_u H_d,
    \label{eq:W_MSSM}
\end{eqnarray}
where $Y_U$, $Y_D$, and $Y_E$ are Yukawa matrices for up-, down-, and
electron-type fermions, respectively, and $\mu_H$ is the SUSY
invariant Higgs mass.  In addition, the soft SUSY breaking terms are
\begin{eqnarray}
    {\cal L}_{\rm soft} &=& 
    -[m_{\tilde{Q}}^2]_{ij} \tilde{Q}_i^* \tilde{Q}_j
    -[m_{\tilde{U}^c}^2]_{ij} \tilde{U}^{c*}_i \tilde{U}^c_j
    -[m_{\tilde{D}^c}^2]_{ij} \tilde{D}^{c*}_i \tilde{D}^c_j
    \nonumber \\ &&
    -[m_{\tilde{L}}^2]_{ij} \tilde{L}_i^* \tilde{L}_j
    -[m_{\tilde{E}^c}^2]_{ij} \tilde{E}^{c*}_i \tilde{E}^c_j
    -m_{H_u}^2 H_u^* H_u
    -m_{H_d}^2 H_d^* H_d
    \nonumber \\ &&
    - (H_u \tilde{U}^c_i [A_{\tilde{U}}]_{ij} \tilde{Q}_j
    - H_d \tilde{D}^c_i [A_{\tilde{D}}]_{ij} \tilde{Q}_j
    - H_d \tilde{E}^c_i [A_{\tilde{E}}]_{ij} \tilde{L}_j + {\rm h.c.})
    - (B_\mu H_u H_d + {\rm h.c.})
    \nonumber \\ &&
    -\frac{1}{2} (
    M_1 \tilde{B} \tilde{B} + M_2 \tilde{W} \tilde{W} 
    + M_3 \tilde{G} \tilde{G} + {\rm h.c.}).
    \label{L_soft}
\end{eqnarray}
In our analysis, we impose the radiative electroweak symmetry breaking
condition; we determine $\mu_H$ and $B_\mu$ parameters so that
$v^2\equiv\langle H_u^0\rangle^2+\langle H_d^0\rangle^2\simeq (174\ 
{\rm GeV})^2$ and $\tan\beta\equiv \langle H_u^0\rangle /\langle
H_d^0\rangle$ are correctly obtained.

In the MSSM, all the soft SUSY breaking parameters given in Eq.\ 
(\ref{L_soft}) are free parameters.  In the framework of the SU(5),
however, that is not the case.  Since $Q$, $U^c$, and $E^c$ ($D^c$ and
$L$) are embedded in ${\bf 10}$ (${\bf \bar{5}}$) representation of
SU(5), soft SUSY breaking parameters for these sfermions should be
unified at the GUT scale.  We parameterize the soft SUSY breaking
parameters at the GUT scale as\footnote
{We assume a simple unification condition for the down-type and
electron-type Yukawa matrices although it does not reproduce realistic
fermion mass texture for the first and second generation fermions.
Our following discussions are, however, insensitive to the parameter
$A_{\tilde{D}}$, and hence this assumption does not change our results
significantly.}
\begin{eqnarray}
    && 
    M_1 = M_2 = M_3 \equiv M_{1/2},
    \label{BC_mg}
    \\ &&
    [m_{\tilde{Q}}^2]_{ij} = 
    [m_{\tilde{U}^c}^2]_{ij} =
    [m_{\tilde{E}^c}^2]_{ij} \equiv 
    m_{\bf 10}^2 \delta_{ij},~~~
    [m_{\tilde{D}^c}^2]_{ij} = 
    [m_{\tilde{L}}^2]_{ij} \equiv 
    m_{\bf \bar{5}}^2 \delta_{ij},
    \\ &&
    m_{H_u}^2 \equiv m_{H{\bf 5}}^2,~~~
    m_{H_d}^2 \equiv m_{H{\bf{\bar{5}}}}^2,
    \\ &&
    A_{\tilde{U}} = a_{\tilde{U}} Y_U,~~~
    A_{\tilde{E}} = A_{\tilde{D}} = a_{\tilde{E}} Y_E.
    \label{BC_A}
\end{eqnarray}
Notice that, in the most general approach, the soft SUSY breaking
masses for the sfermions are not required to be proportional to
$\delta_{ij}$, and sizable off-diagonal elements in the sfermion mass
matrices are possible.  Such off-diagonal elements are, however,
severely constrained since they induce various flavor (and CP)
violating processes like $K^0$-$\bar{K}^0$, $D^0$-$\bar{D}^0$, and
$B^0$-$\bar{B}^0$ mixings, $b\rightarrow s\gamma$, $\mu\rightarrow
e\gamma$, and so on \cite{NPB477-321}.  In addition, in our following
analysis, we focus on the muon MDM and the lightest Higgs mass which
are insensitive to the flavor violations in the sfermion mass
matrices.  Thus, we neglect the effect of the off-diagonal elements in
the following discussions.  In summary, we parameterize the soft SUSY
breaking parameters at the electroweak scale using the following
parameters:
\begin{eqnarray}
    M_{1/2},~~~
    m_{\bf 10}, ~~~
    m_{\bf \bar{5}}, ~~~
    m_{H{\bf 5}}, ~~~
    m_{H{\bf{\bar{5}}}}, ~~~
    a_{\tilde{U}}, ~~~
    a_{\tilde{E}}, ~~~
    \tan\beta,~~~
    {\rm sign}(M_{1/2} \mu_H).
    \label{GUTparams}
\end{eqnarray}
Once these parameters are given, we can calculate the muon MDM and the
lightest Higgs mass as well as the mass spectrum of the
superparticles.

Let us next consider how the muon MDM and the lightest Higgs mass
behave in this framework.  In the MSSM, the supersymmetric
contribution to the muon MDM is from chargino-sneutrino and
neutralino-smuon loop diagrams.  The most important point is that
$a_\mu({\rm SUSY})$ is enhanced when $\tan\beta$ is large.  In the
limit $\tan\beta\gg 1$, the SUSY contribution to the muon MDM is
approximately given by \cite{muonMDM}
\begin{eqnarray}
    a_\mu({\rm SUSY}) &\simeq&
    g_1^2 m_\mu^2 M_1 \mu_H \tan\beta
    \nonumber \\ && \times
    \Bigg[ 
    I^1_5 (M_1^2, M_1^2, m_{\tilde{\mu}L}^2,
    m_{\tilde{\mu}R}^2, m_{\tilde{\mu}R}^2)
    + I^1_5 (M_1^2, M_1^2,m_{\tilde{\mu}L}^2,
    m_{\tilde{\mu}L}^2,m_{\tilde{\mu}R}^2)
    \nonumber \\ &&
    - I^1_5(M_1^2, M_1^2, \mu_H^2,
    m_{\tilde{\mu}R}^2,m_{\tilde{\mu}R}^2)
    - I^1_5(M_1^2, \mu_H^2, \mu_H^2,
    m_{\tilde{\mu}R}^2,m_{\tilde{\mu}R}^2)
    \nonumber \\ &&
    + \frac{1}{2} 
    I^1_5 (M_1^2, M_1^2, \mu_H^2,
    m_{\tilde{\mu}L}^2,m_{\tilde{\mu}L}^2)
    + \frac{1}{2} 
    I^1_5 (M_1^2, \mu_H^2, \mu_H^2,
    m_{\tilde{\mu}L}^2,m_{\tilde{\mu}L}^2)
    \Bigg]
    \nonumber \\ &&
    + g_2^2 m_\mu^2 M_2 \mu_H \tan\beta
    \nonumber \\ && \times
    \Bigg[ 
    -\frac{1}{2}
    I^1_5(M_2^2, M_2^2, \mu_H^2,
    m_{\tilde{\mu}L}^2,m_{\tilde{\mu}L}^2)
    -\frac{1}{2} 
    I^1_5(M_2^2, \mu_H^2, \mu_H^2,
    m_{\tilde{\mu}L}^2,m_{\tilde{\mu}L}^2)
    \nonumber \\ &&
    + 2 I^0_4(M_2^2, M_2^2, \mu_H^2, m_{\tilde{\nu}}^2)
    - I_5^1(M_2^2, M_2^2, \mu_H^2,
    m_{\tilde{\nu}}^2,m_{\tilde{\nu}}^2)
    \nonumber \\ &&
    + 2 I^0_4 (M_2^2, \mu_H^2, \mu_H^2, m_{\tilde{\nu}}^2)
    - I^1_5 (M_2^2, \mu_H^2, \mu_H^2,
    m_{\tilde{\nu}}^2,m_{\tilde{\nu}}^2)
    \Bigg]
\end{eqnarray}
where
\begin{eqnarray}
    I^p_q (m_1^2, \cdots, m_q^2) \equiv \int \frac{d^4 k}{(2\pi)^4i} 
    \frac{(k^2)^p}{(k^2-m_1^2)\cdots (k^2-m_q^2)},
\end{eqnarray}
and $m_{\tilde{\mu}L}^2\equiv [m_{\tilde{L}}^2]_{22}$,
$m_{\tilde{\nu}L}^2\equiv [m_{\tilde{L}}^2]_{22}$, and
$m_{\tilde{\mu}R}^2\equiv [m_{\tilde{E}^c}^2]_{22}$.  For example,
taking $m_{\tilde{\mu}L}^2= m_{\tilde{\mu}R}^2= M_2^2= \mu_H^2\equiv
m_{\rm SUSY}^2$, and neglecting the U(1)$_Y$ contribution, $a_\mu({\rm
SUSY})$ becomes
\begin{eqnarray}
    a_\mu({\rm SUSY}) \simeq \frac{5 g_2^2}{192\pi^2} 
    \frac{m_\mu^2}{m_{\rm SUSY}^2} {\rm sign}(M_2 \mu_H) 
    \tan\beta.
    \label{a_mu(SUSY)_simple}
\end{eqnarray}
When $\tan\beta$ is large, $a_\mu({\rm SUSY})$ can be sizable even if
the superparticles are heavy.  From Eq.\ (\ref{a_mu(SUSY)_simple}),
however, one easily sees that, when $\tan\beta$ is small, (some of)
the superparticles are required to be light so that $a_\mu({\rm
SUSY})$ becomes large enough to explain the BNL E821 anomaly.  In
addition, it is also important to note that $a_\mu({\rm SUSY})$ is
proportional to ${\rm sign}(M_2 \mu_H)$ in the large $\tan\beta$
limit.  Motivated by the E821 anomaly, hereafter, we take ${\rm
sign}(M_2 \mu_H)$ to be positive.

For the case of light superparticles, we must consider various
experimental constraints.  First of all, negative searches for the
superparticles set lower bounds on the masses of the superparticles.
In this letter, as a guideline, we require that all the charged
superparticles be heavier than 100 GeV \cite{PDG}.\footnote
{In the parameter region we are interested in, the first and second
generation squarks as well as gluino are heavier than 300 GeV, and the
experimental constraints on their masses are satisfied.  Thus, we
require $m_{\tilde{t}_1}>100\ {\rm GeV}$ \cite{PDG}.}

In addition, lower bound on the Higgs mass derived by the LEP II
experiment \cite{LEP2Higgs},
\begin{eqnarray}
    m_h\geq 113.5\ {\rm GeV},
    \label{mh_bound}
\end{eqnarray}
provides a severe constraint when $\tan\beta$ is small.  To understand
this fact, it is instructive to see the leading-log formula for the
lightest Higgs mass in the MSSM in the decoupling limit
\cite{Higgs1Loop}:
\begin{eqnarray}
    m_h^2 \simeq m_Z^2 \cos^2 2\beta + 
    \frac{3}{4\pi^2} \frac{m_t^4}{v^2} 
    \log \frac{m_{\tilde{t}}^2}{m_t^2},
    \label{mh}
\end{eqnarray}
where $m_Z$ is the $Z$-boson mass,
$m_{\tilde{t}}\equiv\sqrt{m_{\tilde{t}_1}m_{\tilde{t}_2}}$ is the
geometric mean of the two stop mass eigenvalues $m_{\tilde{t}_1}$ and
$m_{\tilde{t}_2}$, and $m_t$ is the top quark mass.  (Hereafter, we
use $m_t=174.3\ {\rm GeV}$ \cite{PDG} unless otherwise mentioned.)
Here, the first term is the tree-level contribution which becomes
larger when $\tan\beta$ is large.  On the contrary, the second term is
the radiative correction from the top-stop loops, and is enhanced when
the stops become heavier.  Thus, when $\tan\beta$ is small, the stop
masses are required to be heavy to satisfy the constraint
(\ref{mh_bound}).

At this point, it is natural to wonder if the two requirements, one
from the muon MDM and the other from the Higgs mass, can be
simultaneously satisfied when $\tan\beta$ is not large.  To answer
this question, it is crucial to study the mass spectrum and mixings of
the superparticles at the electroweak scale.

In order for precise calculations of physical quantities as functions
of the fundamental parameters listed in (\ref{GUTparams}), we first
calculate the MSSM parameters at the SUSY scale $\mu_{\rm SUSY}$.  For
this purpose, we use the renormalization group equations based on the
standard model for the scale $\mu <\mu_{\rm SUSY}$, and those based on
the MSSM for $\mu_{\rm SUSY}< \mu <M_{\rm GUT}$.  The parameter
$\mu_{\rm SUSY}$ should be regarded as a typical mass scale of the
superparticles; in the following analysis, we take $\mu_{\rm SUSY}$ to
be the geometric mean of the stop masses unless otherwise mentioned.
Then, using the parameters at $\mu=\mu_{\rm SUSY}$, we calculate the
mass spectrum and mixings of the superparticles as well as other
physical quantities.  Using the formula given in \cite{muonMDM}, we
calculate the SUSY contribution to the muon MDM.  In addition, we also
calculate the lightest Higgs boson mass taking account of the dominant
two-loop radiative corrections using {\tt FeynHiggsFast} package
\cite{hph/0002213}.

Before going into the discussion about $a_\mu({\rm SUSY})$ and $m_h$,
we first study the behavior of the sfermion masses (in particular,
stop and smuon masses).  Although the boundary conditions for the soft
SUSY breaking parameters are given in Eqs.\ (\ref{BC_mg}) --
(\ref{BC_A}), the soft SUSY breaking parameters at the electroweak
scale change because of the renormalization group effects.  Taking
$\mu_{\rm SUSY}=500\ {\rm GeV}$ and $\tan\beta=5$, we obtain
\begin{eqnarray}
    m_{\tilde{\mu}_L}^2 (\mu = \mu_{\rm SUSY}) &\simeq& 
    m_{\bf \bar{5}}^2
    + 0.03 m_{H{\bf 5}}^2 - 0.03 m_{H{\bf \bar{5}}}^2 
    + 0.49 M_{1/2}^2,
    \label{fit-smuL}
    \\
    m_{\tilde{\mu}_R}^2 (\mu = \mu_{\rm SUSY}) &\simeq& 
    m_{\bf 10}^2
    - 0.07 m_{H{\bf 5}}^2 + 0.07 m_{H{\bf \bar{5}}}^2 
    + 0.15 M_{1/2}^2,
    \\
    m_{\tilde{t}_L}^2 (\mu = \mu_{\rm SUSY}) &\simeq& 
    0.75 m_{\bf 10}^2 - 0.13 m_{H{\bf 5}}^2 
    + 0.01 m_{H{\bf \bar{5}}}^2
    + 4.12 M_{1/2}^2
    \nonumber \\ &&
    - 0.11 a_{\tilde{U}} M_{1/2} 
    - 0.03 a_{\tilde{U}}^2,
    \\
    m_{\tilde{t}_R}^2 (\mu = \mu_{\rm SUSY}) &\simeq& 
    0.51 m_{\bf 10}^2 - 0.20 m_{H{\bf 5}}^2 
    - 0.04 m_{H{\bf \bar{5}}}^2
    + 2.94 M_{1/2}^2 
    \nonumber \\ &&
    - 0.23 a_{\tilde{U}} M_{1/2} 
    - 0.06 a_{\tilde{U}}^2,
\end{eqnarray}
where $m_{\tilde{t}_L}^2\equiv [m_{\tilde{Q}}^2]_{33}$, and
$m_{\tilde{t}_R}^2\equiv [m_{\tilde{U}^c}^2]_{33}$.  From these
relations, we expect rich sparticle mass spectrum. This model has a
significant contrast with the CMSSM where the universal scalar mass is
assumed: $m_{\bf 10}^2=m_{\bf \bar{5}}^2=m_{H{\bf
5}}^2=m_{H{\bf{\bar{5}}}}^2\equiv m_0^2$.  In the CMSSM, all the
sfermion masses have strong correlations because all the sfermion
masses are increased (decreased) if we adopt larger (smaller) values
of $m_0$ and/or $M_{1/2}$.  Thus, in the CMSSM, it is difficult to
explain the E821 anomaly without conflicting the Higgs mass constraint
if $\tan\beta$ is not large.  In the general SUSY SU(5) model,
however, this is not the case since the correlation among the sfermion
masses becomes weak.  This fact has a very important implication as we
will see below.

\begin{figure}
    \centerline{\epsfxsize=0.75\textwidth\epsfbox{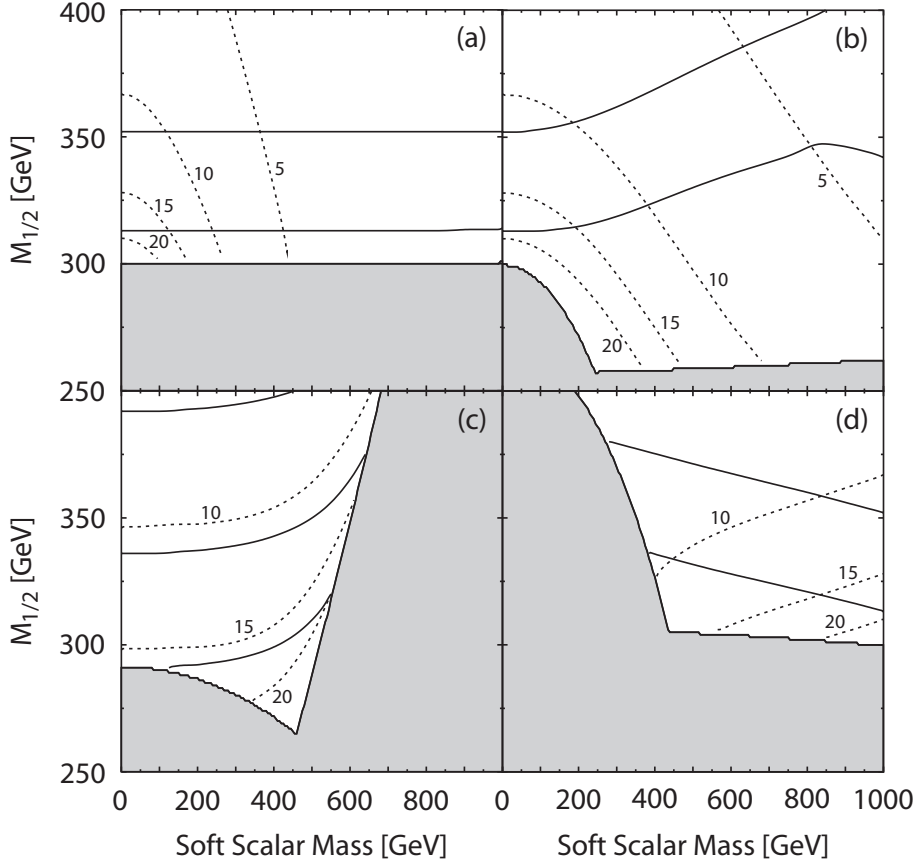}}
    \caption{Contours of constant $a_\mu({\rm SUSY})$ (dotted).
    (Values of $a_\mu({\rm SUSY})$ are shown in the figures in units
    of $10^{-10}$. The vertical axis is $M_{1/2}$, and the horizontal
    axis is (a) $m_{\bf \bar{5}}$, (b) $m_{\bf 10}$, (c) $m_{H{\bf
    5}}$, and (d) $m_{H{\bf \bar{5}}}$.  Here, we take $\tan\beta=5$,
    and $a_{\tilde{U}}=a_{\tilde{E}}=0$.  In addition, other
    parameters are (a) $m_{\bf 10}=0$, $m_{H{\bf 5}}=520\ {\rm GeV}$,
    $m_{H{\bf \bar{5}}}=1\ {\rm TeV}$, (b) $m_{\bf \bar{5}}=0$,
    $m_{H{\bf 5}}=520\ {\rm GeV}$, $m_{H{\bf \bar{5}}}=1\ {\rm TeV}$,
    (c) $m_{\bf \bar{5}}=0$, $m_{\bf 10}=0$, $m_{H{\bf \bar{5}}}=1\ 
    {\rm TeV}$, and (d) $m_{\bf \bar{5}}=0$, $m_{\bf 10}=0$, $m_{H{\bf
    5}}=520\ {\rm GeV}$.  The shaded region is excluded by the
    negative searches for charged superparticles.  Contours of the
    constant $m_h$ are also shown in the solid lines ($m_h=110\ {\rm
    GeV}$, 111 GeV (and 112 GeV for (c)) from below).}
    \label{fig:amumh}
\end{figure}

Now, we are at the position to discuss the SUSY contribution to the
muon MDM as well as the lightest Higgs mass.  In fact, $m_h$ is
sensitive to the value of $a_{\tilde{U}}$.  Therefore, we split our
discussion into two cases: one with relatively small $a_{\tilde{U}}$
and the other with large $a_{\tilde{U}}$.

We start our discussion with the former case.  In Fig.\ 
\ref{fig:amumh}, we plot contours of the constant $a_\mu({\rm SUSY})$
and $m_h$ on $m_{\bf \bar{5}}$ vs.\ $M_{1/2}$, $m_{\bf 10}$ vs.\ 
$M_{1/2}$, $m_{H{\bf 5}}$ vs.\ $M_{1/2}$, and $m_{H{\bf \bar{5}}}$
vs.\ $M_{1/2}$ planes, with $a_{\tilde{U}}=0$.  Notice that, in the
parameter region we discuss below, we checked that the LSP is the
neutral superparticles (the lightest neutralino or the sneutrino).

First, we discuss behaviors of $a_\mu({\rm SUSY})$.  For this purpose,
let us point out that the dominant contribution to $a_\mu({\rm SUSY})$
is from the chargino-sneutrino diagram.  Consequently, $a_\mu({\rm
SUSY})$ is more enhanced with lighter charginos and lighter
left-handed sleptons.  Based on this fact, dependence on $M_{1/2}$ can
be understood; for larger value of $M_{1/2}$, heavier superparticles
are realized, resulting in suppressed $a_\mu({\rm SUSY})$.  In
addition, $m_{\bf \bar{5}}$ dependence is also trivial; with larger
value of $m_{\bf \bar{5}}$, the left-handed slepton masses become
larger and hence $a_\mu({\rm SUSY})$ becomes smaller. (See Fig.\ 
\ref{fig:amumh}a.)  Slight dependence on $m_{H{\bf \bar{5}}}$ is from
the renormalization group effect on $m_{\tilde{\mu}_L}$.  As can be
seen in Eq.\ (\ref{fit-smuL}), $m_{H{\bf \bar{5}}}$ gives a negative
contribution to $m_{\tilde{\mu}_L}^2$.  Thus, to obtain a larger value
of $a_\mu({\rm SUSY})$, $m_{H{\bf \bar{5}}}$ should be increased.
(See Fig.\ \ref{fig:amumh}d.)  Dependences on $m_{\bf 10}$ and
$m_{H{\bf 5}}$ arise since the $\mu_H$ parameter determined by the
radiative electroweak symmetry breaking condition depends on these
parameters; $\mu_H$ increases for larger value of $m_{\bf 10}$ and for
smaller value of $m_{H{\bf 5}}$.  Since $\mu_H$ (almost) corresponds
to the Higgsino-like chargino mass, larger value of $\mu_H$ gives rise
to smaller value of $a_\mu({\rm SUSY})$.  This results in the
behaviors of $a_\mu({\rm SUSY})$ shown in Figs.\ \ref{fig:amumh}b and
\ref{fig:amumh}c.

Now, let us consider the Higgs mass.  The Higgs mass is sensitive to
$M_{1/2}$.  This is because, in the parameter region given in the
figures, stop masses are primarily determined by $M_{1/2}$.
Importantly, as $M_{1/2}$ increases, the stop masses are more
enhanced, resulting in larger value of $m_h$.  On the contrary, the
Higgs mass is relatively insensitive to the scalar masses.

\begin{figure}[t]
    \centerline{\epsfxsize=0.75\textwidth\epsfbox{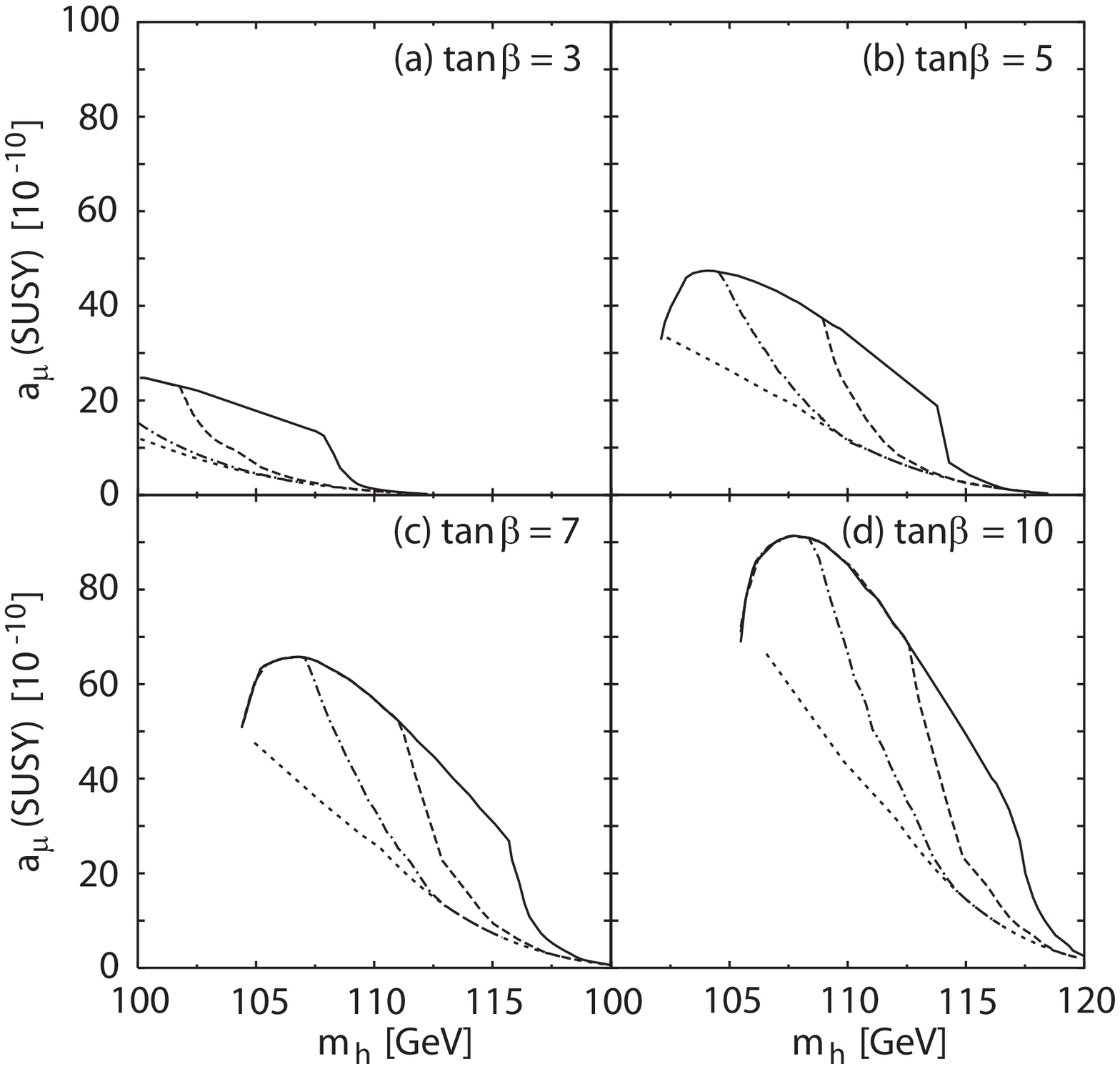}}
    \caption{Maximum possible value of $a_\mu({\rm SUSY})$ in units of 
    $10^{-10}$ as a function of the lightest Higgs mass $m_h$.  
    We take $m_{\rm max}=500\ {\rm GeV}$ (dot-dashed), 1 TeV (dashed),
    and 2 TeV (solid), $a_{\tilde{U}}=a_{\tilde{E}}=0$, and (a)
    $\tan\beta=3$, (b) $\tan\beta=5$, (c) $\tan\beta=7$, and (d)
    $\tan\beta=10$.  The case with the CMSSM is also shown in the
    dotted lines.}
    \label{fig:amumax}
\end{figure}

It is interesting to plot the maximum possible value of $a_{\mu}({\rm
SUSY})$ as a function of the lightest Higgs mass.  For this purpose,
we vary the parameters $m_{\bf \bar{5}}$, $m_{\bf 10}$, $m_{H{\bf
5}}$, $m_{H{\bf \bar{5}}}$, and $M_{1/2}$ from 0 to $m_{\rm max}$,
where we take $m_{\rm max}$ to be 500 GeV, 1 TeV, and 2 TeV, and
obtain the upper bound on $a_{\mu}({\rm SUSY})$ for a given value of
$m_h$.  The results are shown in Fig.\ \ref{fig:amumax}.  As $m_h$
increases, the maximum possible value of $a_{\mu}({\rm SUSY})$
decreases.  This is because, to obtain a larger value of $m_h$,
$M_{1/2}$ is required to be large to enhance the radiative correction
by pushing up the stop masses through the running effects.  As a
result, other sparticle masses are also suppressed for larger value of
$m_h$ and the upper bound becomes smaller.  In addition, we can find a
``kink'' on each plot.  This is from the fact that, to obtain the
maximum possible value of $a_{\mu}({\rm SUSY})$, $m_{H{\bf \bar{5}}}$
is preferred to be large.  However, when $M_{1/2}$ is small, the stau
becomes lighter than the experimental bound if $m_{H{\bf \bar{5}}}$ is
too large.  Therefore, as $m_h$ is reduced, the upper bound on
$m_{H{\bf \bar{5}}}$ becomes smaller than $m_{\rm max}$.  On the
contrary, when $m_h$ is large enough, $m_{H{\bf \bar{5}}}=m_{\rm max}$
is allowed.  The kink corresponds to the boundary of these two
parameter regions.  If we change $m_{\rm max}$, the upper bound on
$a_{\mu}({\rm SUSY})$ also changes; when $m_h$ is large, the maximum
possible value of $a_{\mu}({\rm SUSY})$ increases by adopting larger
value of $m_{\rm max}$.

From Fig.\ \ref{fig:amumax}, we see that the negative search for the
Higgs boson at LEP II places a severe constraint on the possible value
of the SUSY contribution to the muon MDM for small $\tan\beta$ case.
For $\tan\beta\lesssim 5$, $a_{\mu}({\rm SUSY})$ cannot explain the
E821 anomaly even at the 2-$\sigma$ level if we adopt $m_{\rm max}=1\ 
{\rm TeV}$.  Of course, with larger value of $\tan\beta$,
$a_{\mu}({\rm SUSY})$ may become larger and it is possible to explain
the E821 anomaly.

We should note here that the result depends on the top quark mass,
since the radiative correction to the lightest Higgs mass is sensitive
to $m_t$.  The lightest Higgs mass is enhanced for larger value of
$m_t$.  Therefore, for a given value of $m_h$, we can push up the
maximum possible value of $a_{\mu}({\rm SUSY})$ by increasing $m_t$.
We checked that, if we use $m_t=179.4\ {\rm GeV}$ which is the
1-$\sigma$ upper bound on the top quark mass \cite{PDG}, the curves
move to the right; approximately, the same upper bound on
$a_{\mu}({\rm SUSY})$ is obtained for the Higgs mass larger than about
2 $-$ 3 GeV compared to the previous case ($m_t=174.3\ {\rm GeV}$).

We can also compare our results with those with the CMSSM.  To
maximize $a_{\mu}({\rm SUSY})$ in the CMSSM framework, we repeat our
analysis imposing $m_{\bf 10}=m_{\bf \bar{5}}=m_{H{\bf
5}}=m_{H{\bf{\bar{5}}}}$.  We found that the result for the CMSSM is
independent of $m_{\rm max}$ as far as $m_{\rm max}\geq 500\ {\rm
GeV}$.  The results are also shown in Fig.\ \ref{fig:amumax} in the
dashed lines.  As one can see, the maximum possible value for the
CMSSM case is significantly smaller than that in the general SU(5)
case since the number of the free parameters is much smaller.  In
particular, $\tan\beta\gtrsim 10$ is required in the CMSSM case to
explain the E821 anomaly while $\tan\beta\gtrsim 7$ in the general
SU(5) GUT approach for $m_{\rm max}=1\ {\rm TeV}$.

Now, we consider the second case with large $a_{\tilde{U}}$.\footnote
{If $a_{\tilde{U}}$ is large, color breaking minimum may exist and the
origin of the squark potential may become a false vacuum.  Such a
situation is, however, cosmologically safe if the squark field is
trapped in the false vacuum from in early universe.  For example,
thermal effect in the early universe can trap the squark field at the
origin.}
In this case, the Higgs mass may be affected by the
large trilinear coupling.  To understand this fact, it is instructive
to calculate the correction to the quartic coupling of the
standard-model like Higgs boson which is approximately given by
$H_{\rm SM}\simeq H_u\sin\beta+H_d\cos\beta$.  Denoting the potential
of $H_{\rm SM}$ below the SUSY scale as $V=\frac{1}{2}\lambda(|H_{\rm
SM}|^2-v^2)^2$, we obtain, at the tree level,
$\lambda=\frac{1}{4}(g_2^2+g_1^2)\cos^22\beta$ with $g_2$ and $g_1$
being the gauge coupling constants for SU(2)$_L$ and U(1)$_Y$ gauge
interactions, respectively.  If the trilinear coupling is large,
however, the threshold correction to $\lambda$ becomes sizable.
Assuming a large hierarchy between the electroweak scale and the stop
mass, and approximating $m_{\tilde{t}_1}\sim m_{\tilde{t}_2}$,
the threshold correction to the quartic coupling from the stop
loop is given by \cite{PLB262-54}
\begin{eqnarray}
    \Delta\lambda \simeq \frac{3}{8\pi^2}
    \left( \frac{y_t^2 A_{\tilde{t}}^2}{m_{\tilde{t}}^2}
        - \frac{1}{12} \frac{A_{\tilde{t}}^4}{m_{\tilde{t}}^4}
    \right) \sin^4\beta.
\end{eqnarray}
Here $y_t=[Y_U]_{33}$ and $A_{\tilde{t}}=[A_{\tilde{U}}]_{33}$, and 
the fitting formula for $A_{\tilde{t}}$ is given by
\begin{eqnarray}
    A_{\tilde{t}} (\mu = \mu_{\rm SUSY}) \simeq
    1.70 M_{1/2} + 0.24 a_{\tilde{U}},
    \label{A_t}
\end{eqnarray}
where we used $\mu_{\rm SUSY}=500\ {\rm GeV}$ and $\tan\beta=5$.
Notice that $\Delta\lambda$ stays finite even if $m_{\tilde{t}}$
increases as far as the ratio $A_{\tilde{t}}/m_{\tilde{t}}$ is fixed.
When $a_{\tilde{U}}=0$ (or $a_{\tilde{U}}$ is small), $A_{\tilde{t}}$
is small and hence the trilinear coupling does not affect the Higgs
mass so much.  If a large value of $a_{\tilde{U}}$ is adopted,
however, $A_{\tilde{t}}$ is enhanced and $\Delta\lambda$ can be close
to $\sim 0.1$.  In this case, $m_h$ is drastically enhanced even if
the stops are relatively light.

Of course, $a_{\tilde{U}}$ also affects other parameters.  One
important effect is that the $\mu_H$ parameter increases as
$A_{\tilde{t}}$ increases.  This is because the trilinear coupling
changes the value of $m_{H_u}^2$ through the renormalization group
effect.  As a result, too large $a_{\tilde{U}}$ results in a
suppressed value of $a_\mu({\rm SUSY})$.

\begin{figure}[t]
    \centerline{\epsfxsize=0.5\textwidth\epsfbox{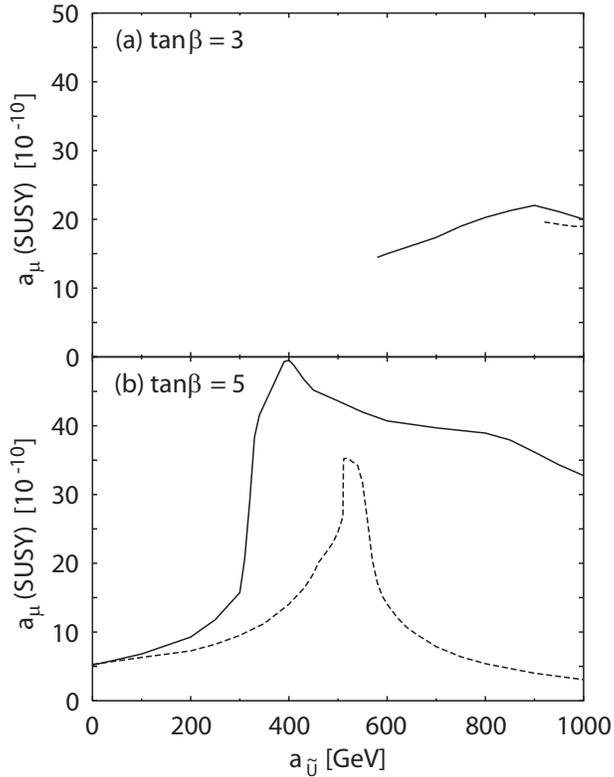}}
    \caption{Maximum possible value of $a_\mu({\rm SUSY})$ in units of 
    $10^{-10}$ as a function of $a_{\tilde{U}}$ for $\tan\beta=3$ and
    $5$.  The solid lines are for the general SU(5) case and the
    dashed lines are for the CMSSM case, and we adopt $m_{\rm max}=1\ 
    {\rm TeV}$.  The lightest Higgs mass $m_h$ is fixed to be 113.5
    GeV.  Notice that, for $\tan\beta=3$, $a_{\tilde{U}}$ has to be
    large enough to push up the Higgs mass, and hence the lower bound
    on $a_{\tilde{U}}$ exists.}
    \label{fig:a-dep}
\end{figure}

To study the effect of $a_{\tilde{U}}$, we vary $a_{\tilde{U}}$ as
well as other soft parameters and obtain the maximum possible value of
$a_\mu({\rm SUSY})$ as a function of $m_h$.  In Fig.\ \ref{fig:a-dep},
we plot the maximum possible value as a function of $a_{\tilde{U}}$
for $m_h=113.5\ {\rm GeV}$.  For the $\tan\beta=5$ case, by assuming a
large value of $a_{\tilde{U}}$, the SUSY contribution to the muon MDM
is significantly enhanced relative to the case of $a_{\tilde{U}}=0$.
In the general SU(5) case, we find a big increase of the upper bound
on $a_\mu({\rm SUSY})$ at around $a_{\tilde{U}}\sim 300\ {\rm GeV}$.
This can be understood as follows.  When $a_{\tilde{U}}$ is large, the
lightest Higgs mass can be enhanced by a large value of
$\Delta\lambda$ without pushing up the sfermion masses.  As a result,
the slepton masses and the chargino masses may be small even for a
large value of $m_h$, and hence the SUSY contribution to the muon MDM
may become large.  We see that $a_\mu({\rm SUSY})$ can be as large as
the deviation between $a_{\mu}({\rm E821})$ and $a_{\mu}({\rm SM})$
even with a small value of $\tan\beta$, like $\tan\beta=5$.  In
addition, the trilinear coupling may also play a significant role in
the case of the CMSSM, as can be seen in Fig.\ \ref{fig:a-dep}.

In summary, we have discussed the muon magnetic dipole moment and the
Higgs mass in the framework of the supersymmetric SU(5) models.
Importantly, we have {\sl not} adopted the assumption of the universal
scalar mass but have treated all the relevant parameters to be free.
Then, we found that the maximum possible value of the SUSY
contribution to the muon MDM becomes larger compared to the case of
the universal scalar mass.  When the trilinear coupling is small, to
maximize $a_\mu({\rm SUSY})$ for a fixed value of $m_h$, soft SUSY
breaking masses for the sfermions at the GUT scale should be small
while those for the Higgses (as well as the gaugino masses) are
preferred to be finite.  It is interesting that such a situation may
be realized in, for example, the gaugino-mediated SUSY breaking
scenario with the Higgs multiplets in the bulk \cite{gaugino-med}.  In
such a framework, the gauge and Higgs multiplets directly feel the
effect of the SUSY breaking, and hence $M_{1/2}$, $m_{H{\bf{5}}}$, and
$m_{H{\bf{\bar{5}}}}$ are finite while $m_{\bf 10}$ and $m_{\bf
\bar{5}}$ vanish at the cutoff scale.  In addition, it has been also
shown that, if the trilinear scalar coupling for the stop is large,
constraint from the lightest Higgs mass is drastically relaxed.  In
this case, the SUSY contribution to the muon MDM can completely
explain the E821 anomaly even for $\tan\beta=5$.

{\em Acknowledgment:} One of the authors (M.E.) thanks N. Abe and
S. Komine for useful discussions.  This work is supported by the
Grant-in-Aid for Scientific Research from the Ministry of Education,
Science, Sports, and Culture of Japan, No.\ 12047201 and No.\
13740138.

\end{document}